\documentclass[twocolumn,prb,superscriptaddress,showpacs,citeautoscript,floatfix]{revtex4}
\usepackage{graphicx}

\begin{document}


\title{Raman scattering in osmium under pressure}

\date{\today}

\author{Yu.\ S. Ponosov}

\affiliation{Max-Planck-Institut f{\"u}r Festk{\"o}rperforschung,
Heisenbergstra{\ss}e 1, D-70569 Stuttgart, Germany}

\affiliation{Institute of Metal Physics UD RAS, 620219, S. Kovalevskaya
str.\ 18, Ekaterinburg, Russia}

\author{I. Loa}

\email[E-mail: ]{I.Loa@fkf.mpg.de}

\affiliation{Max-Planck-Institut f{\"u}r Festk{\"o}rperforschung,
Heisenbergstra{\ss}e 1, D-70569 Stuttgart, Germany}

\author{V. E. Mogilenskikh}

\affiliation{Institute of Metal Physics UD RAS, 620219, S. Kovalevskaya
str.\ 18, Ekaterinburg, Russia}

\author{K. Syassen}

\affiliation{Max-Planck-Institut f{\"u}r Festk{\"o}rperforschung,
Heisenbergstra{\ss}e 1, D-70569 Stuttgart, Germany}

\begin{abstract}
The effect of pressure and temperature on the Raman-active phonon mode of
osmium metal has been investigated for pressures up to 20~GPa and temperatures
in the range 10--300~K. Under hydrostatic conditions (He pressure medium) the
phonon frequency increases at a rate of 0.73(5)~cm$^{-1}$/GPa ($T=300$~K). A
large temperature-induced and wavelength-dependent frequency shift of the
phonon frequency is observed, of which only a small fraction can be associated
with the thermal volume change. The main contribution to the temperature
dependence of the phonon frequency is rather attributed to non-adiabatic
effects in the electron-phonon interaction, which explains also the
observation of an increasing phonon line width upon cooling. The phonon line
width and the pressure-induced frequency shift were found to be unusually
sensitive to shear stress.
\end{abstract}

\smallskip


\maketitle


Laser Raman scattering (RS) has been utilized in recent diamond anvil cell
studies of the lattice dynamical properties of elemental metals at high
pressures~\cite{Olijnyk,Goncharov}. In metals, phonon RS is essentially
restricted to first-order scattering by zone-center optical phonon modes
because of the relatively weak Raman signals. Therefore, it is mainly the
elemental hcp metals (having one first-order Raman-active phonon mode) that
have been investigated by high-pressure RS with a focus on phonon frequency
changes which yield indirect information the pressure/volume dependence of the
elastic constant~$C_{44}$.

The high resolution of RS also allows for an investigation of the
electron-phonon interaction which is of specific importance to the lattice
dynamics of metals. Motivated by the early Raman studies on elemental
metals~\cite{0}, the effect of electron-phonon interaction on the
long-wavelength optical phonon spectrum was studied theoretically~\cite{6}.
The central result is a prediction of a large \textit{non-adiabatic}
renormalization of the phonon spectrum and a strong dispersion of the optical
phonon branch at very small phonon momenta, $q \lesssim \hbar\omega_{0} /
v_F$, where $\omega_{0}$ is the bare phonon frequency and $v_F$ the Fermi
velocity. At these small momenta, the phonon phase velocity is larger than the
electronic one, which leads to a violation of the adiabatic approximation and
disappearance of the electron-hole channels for the phonon decay.

Most RS studies of elemental metals were conducted at room temperature
where the effects of electron-phonon interaction vanish, partly due to
the smearing of electronic self-energy effects which depend on details
of the electronic structure. There are only a few ambient-pressure
experimental studies of the temperature and momentum dependences of the
non-adiabatic effects in hcp transition metals~\cite{1,2,3,4}. These
investigations revealed anomalies in the temperature and momentum
dependences of the energy and line width of long-wavelength optical
phonons. In particular, upon cooling they evidence an unusually large
phonon frequency hardening and an anomalous \textit{increase} of the
phonon line width.

Osmium is one of the better-investigated examples. Here, a large and strongly
anisotropic dispersion $d \omega/d q$ of the phonon frequency at small wave
vectors $q \approx 10^{6}$~cm$^{-1}$ was reported~\cite{1}. For transverse
optical phonons it is larger than 10$^{6}$~cm/s and exceeds the typical
dispersion of optical phonon branches in metals by two orders of magnitude.
This result is clearly reminiscent of the theoretical predictions for the case
of non-adiabatic electron-phonon interaction~\cite{5,6,7}. Non-adiabatic
processes were therefore proposed as the key effect to interpret the
experimental observations for the hcp transition metals~\cite{1}, but their
significance was questioned in other studies~\cite{8,9,10,11}. A crucial point
is whether the large phonon frequency increase upon cooling originates simply
from the common anharmonic behavior (i.e., volume change as a function of
temperature) or whether a non-adiabatic electron-phonon interaction plays the
dominant role. A high-pressure experiment offers the possibility to answer
this question by determining the isothermal volume dependence of the relevant
phonon frequencies.

\begin{figure*}[t]
\centerline{\includegraphics[width=0.9\hsize]{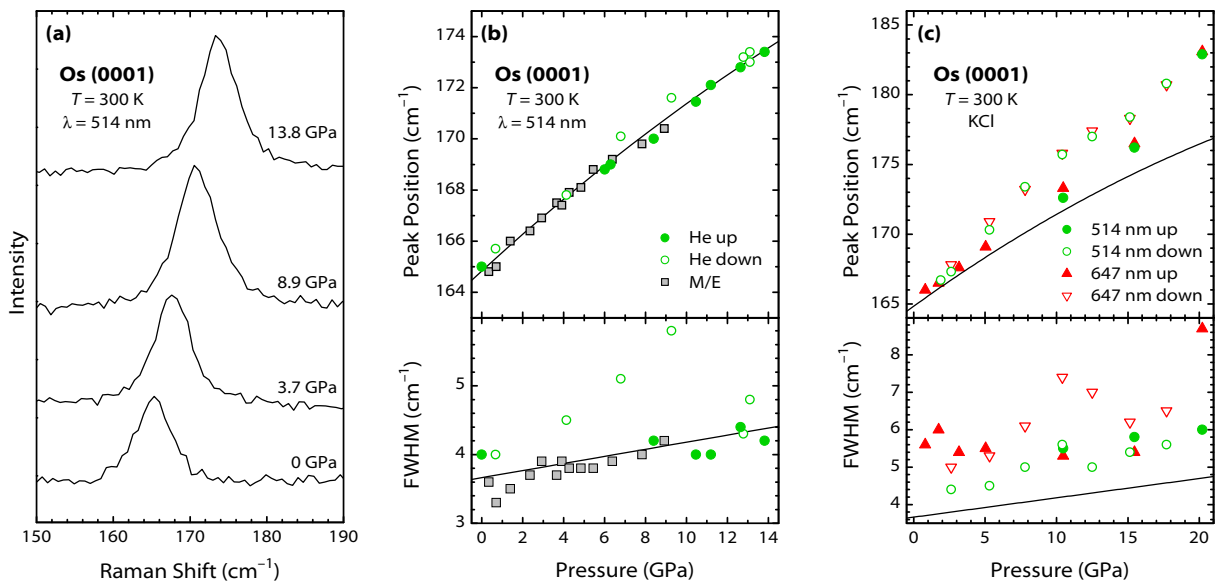}}
\caption{\label{fig1} (Color online) (a) Room-temperature Raman spectra
of osmium at selected pressures, measured in $z(xy)\overline{z}$ polarization.
Helium was used as a pressure-transmitting medium and the spectra were excited
at 514~nm. (b) Raman line frequency and Lorentzian line width (see text) of
osmium as a function of pressure ($T=300$~K, $\lambda=514$~nm). Data were
measured for both up- and down-stroke. A few data points recorded with a
methanol/ethanol medium (up to 9~GPa) are included. (c) Same as (b) but for a
KCl pressure medium. Two excitation wavelengths, 514 and 647~nm, were used.
Lines represent the interpolated results from the runs with He and
methanol/ethanol pressure media, cf.~(b).}
\end{figure*}

The non-adiabatic effects are expected to be influenced by the
application of pressure, through a modification of the electronic band
structure near the Fermi level. Besides, ambient-pressure
results~\cite{1,2,3,4} imply that the main channel of the optical
phonon damping in transition metals is a decay into electron-hole pairs
rather than into a continuum of low energy phonons. This situation may
also change under pressure, leading to a modification of the
temperature behavior of the phonon self-energies. Therefore,
high-pressure Raman measurements need to be performed in a wide
temperature range from ambient down to liquid-helium temperatures in
order to observe clearly the modification of the Raman spectra due to
low-energy electronic excitations and associated changes in the phonon
self-energy.

In this work we report the effect of pressure (up to 20~GPa) \textit{and}
temperature (10--300~K) on the Raman response of osmium metal. The electronic
structure of osmium ($5d^{6} 6s^{2}$) is characterized by a multi-sheet Fermi
surface~\cite{25,26a,26b}. The optical properties were studied
experimentally~\cite{27} and discussed on the basis of electronic band
structure calculations~\cite{26a,26b}. The lattice dynamics was studied at
ambient pressure by Raman~\cite{1,2} and micro-contact spectroscopy~\cite{28}.
The mode Gr{\"u}neisen parameter of the Raman-active phonon was estimated in a
recent pressure experiment up to 4~GPa~\cite{23}. Osmium exhibits a very low
compressibility \cite{29,30,31}, only 10\% larger than that of diamond. The
central aim of our study is to assess the relative importance of anharmonicity
versus non-adiabatic electron-phonon interaction. The experimental results are
used to test the conclusions of previous ambient-pressure Raman studies
regarding the importance of non-adiabatic effects.


Plates of osmium single crystals with [0001] orientation were thinned to
20~$\mu$m thickness by a combination of mechanical and electro-polishing. The
high purity of the crystals (``clean limit regime'') is indicated by the
resistivity ratio $\rho_{300\,\rm{K}}/
\rho_{4.2\,\rm{K}} > 1000$. The samples were loaded into a diamond anvil
cell and placed in an optical cryostat covering the 10--300~K
temperature range. Three different pressure-transmitting media were
used in the experiments: helium, a 4:1 methanol-ethanol (M/E) mixture,
and KCl. Helium was used as a pressure medium to provide hydrostatic
conditions at low temperatures. Pressures were determined by the ruby
luminescence method~\cite{32} with correction for the temperature
dependence of the ruby emission wavelength~\cite{33}. Raman spectra
were recorded in quasi-backscattering geometry using a triple-grating
spectrometer (Jobin Yvon T64000) equipped with a liquid-nitrogen-cooled
CCD detector. Argon (514~nm) and krypton (647~nm) ion laser lines at
powers of 40 mW were used for excitation. The focal spot on the sample
was about 50 $\mu$m in diameter. A spectrometer bandwidth of
2.5~cm$^{-1}$ was chosen as the optimum compromise between spectral
resolution and Raman intensity.


Figure \ref{fig1}(a) shows selected high-pressure Raman spectra of a single
crystal of osmium measured at room temperature using the 514-nm excitation and
helium as the pressure-transmitting medium. A single symmetric Raman mode is
observed (for the polarization configurations $z(xx)\overline{z}$ and
$z(xy)\overline{z}$) which shifts to higher frequencies with increasing
pressure. The \textit{E}$_{2g}$ vibration corresponds to an anti-phase
displacement (in the basal plane) of the two atoms of the unit cell. In order
to determine accurately the peak position and phonon line width, the spectra
were fitted using a Voigt peak profile, i.e.\ a Lorentzian convoluted with a
Gaussian, where the width of the latter corresponded to the spectral
resolution (2.5~cm$^{-1}$).

Figure \ref{fig1}(b) shows the phonon frequency and line width as a
function of pressure, obtained in several runs at room temperature and
using the 514-nm excitation. The data recorded with the M/E pressure
medium (up to 9 GPa) are consistent with the results using helium. Some
measurements with excitation at 647~nm gave (within the experimental
uncertainty) the same frequencies; the line widths were approximately
1~cm$^{-1}$ larger at the highest pressure.

The pressure dependence of the Raman mode frequency is described well
by a second order polynomial $ \omega(P) = \omega_0 + \alpha P + \beta
P^2$ with $\omega_{0}$ = 164.9(1) cm$^{-1}$, $\alpha$ =
0.73(5)~cm$^{-1}$/GPa, and $\beta$ = -- 0.008(3)~cm$^{-1}$/GPa$^{2}$.
The frequency at zero pressure agrees with earlier results obtained
with the same laser wavelength~\cite{1,2}. From the initial rate
$\alpha = (d \omega/d P)_{P=0}$ and a bulk modulus value of
$B_0$=~400(10)~GPa~\cite{30,31} we determined the mode Gr{\"u}neisen
parameter $\gamma =-\partial \ln \omega /\partial \ln V = 1.77(12)$.
This value is significantly larger than estimated from a previous
experiment up to 4~GPa~\cite{23}. The phonon line width increases by
less than 1~cm$^{-1}$ up to the highest pressure of 14~GPa. In the run
with helium pressure medium we noticed an increased phonon line width
when the pressure was released below 9~GPa. The origin of the
broadening remains uncertain at this point.

The pressure dependences of phonon energy and line width obtained when
using the KCl pressure medium are presented in Fig.~\ref{fig1}(c)
together with the fitted curves from Fig.~\ref{fig1}(b) for the He and
M/E runs. When using KCl, a solid but relatively soft medium, the
phonon frequency increases sharply above 15~GPa and shows kinks near 15
and 5~GPa for the down-stroke. The phonon line widths are larger by
$\sim$25\% than in the He run, even at low pressures. These effects
appear to result from non-hydrostatic strain and indicate a high
sensitivity of the osmium properties to shear stress. It should be
mentioned that no such effect was observed for rhenium and ruthenium,
which were studied to much higher pressures without using any pressure
transmitting medium~\cite{Olijnyk,Goncharov}.

\begin{figure}[t]
\includegraphics[width=\hsize]{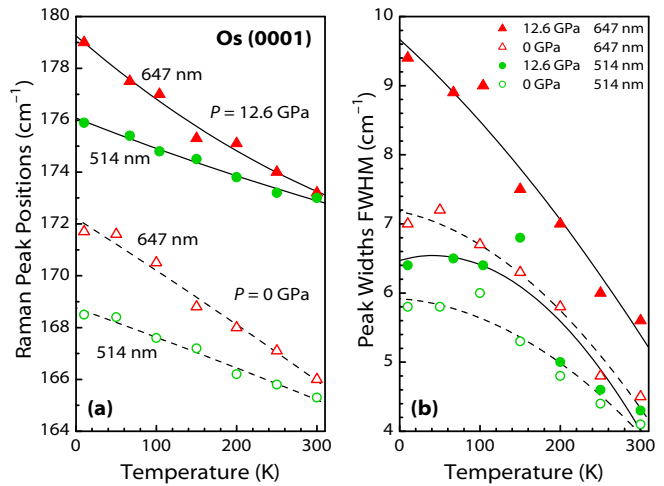}
\caption{\label{fig2} (Color online) (a) Raman peak position and (b) Raman line
width of osmium as a function of temperature measured at 0 and 12.6~GPa with
514~nm (circles) and 647~nm (triangles) excitation wavelengths. Helium was
used as pressure-transmitting medium. Lines are guides to the eye.}
\end{figure}

The temperature dependences of the phonon energy and line width are shown in
Fig.~\ref{fig2}(a) for two pressures and two excitation wavelengths (514 and
647~nm). A strong dependence of the phonon frequency on the excitation energy
is evident from Fig.~\ref{fig2}(a), in agreement with previous work~\cite{1}.
The result was interpreted as a manifestation of a strong phonon dispersion
\textit{vs} probed wave vector \textit{q}. (One should note here that the
variation of \textit{q} with the wavelength $\lambda$ of the incident
radiation is larger than $q\propto 1/\lambda $ due to the wavelength
dependence of the optical properties of the metal~\cite{27}.) The spectra
taken with 514-nm and 647-nm excitation probe wave vector distributions with
$q_{max}$ equal to $1.64 \times 10^{6}$~cm$^{-1}$ and $1.01
\times 10^6$~cm$^{-1}$, respectively. The previous ambient-pressure
experiments on osmium~\cite{1} indicated that in the first case (514~nm) the
incident radiation probes the adiabatic range of momenta, $q
> q_0 \equiv \hbar \omega_0 / v_F$ while in the second case
(647~nm) one probes excitations near $q_0$.

Figure~\ref{fig2}(a) demonstrates that the application of pressure has
only little effect on the phonon dispersion (i.e., dependence on
excitation wavelength at low temperatures) and the temperature
dependence of the phonon energies. The temperature-induced shifts
between 300 and 10~K amount to 3--4~cm$^{-1}$ for excitation at 514~nm
and to 6--7~cm$^{-1}$ for excitation at 647~nm. These values can be
compared with the expected phonon frequency shift due to thermal
expansion. The volume change between 300~K and 10~K amounts to less
than 0.4\%~\cite{34}. With the mode Gr{\"u}neisen parameter $\gamma
=1.77(12)$  determined above we arrive at an expected frequency
increase of 1.3~cm$^{-1}$ due to thermal contraction between 300 and
10~K. In contrast, the actually observed temperature-induced shifts are
larger by factors of 3 and 5 for excitation at 514 and 647~nm,
respectively. Thus, only a small part of the temperature-induced shift
can be attributed to the change in volume.

The unusual increase in phonon line width with decreasing temperature and the
pronounced dependence on the excitation wavelength persist under pressure
[Fig.~\ref{fig2}(b)]. It suggests that the contribution of multi-phonon decay
to the phonon line width is negligible (even if modified under pressure).
Altogether, the experimental results presented here yield direct evidence that
the usual anharmonic processes are not the main cause of the large phonon
hardening upon cooling. They rather support the picture that non-adiabatic
effects in the electron-phonon interaction play a dominant role and lead to
the observed anomalies in the phonon self-energies.

Finally, we consider a possible relationship between non-adiabatic
electron-phonon interaction and a high sensitivity to non-hydrostatic stresses
of the phonon self-energies in osmium. The observation of increased line
widths in the spectra measured at 12.6~GPa [especially with 647-nm excitation,
cf.~Fig.~\ref{fig2}(b)] is an indication of the non-adiabatic effects being
influenced by \textit{hydrostatic} pressure, even at room temperature. The
application of pressure thus seems to affect those details of the electronic
band structure near the Fermi level that cause the non-adiabatic effects. In
the case of \textit{non-hydrostatic} compression, the electronic band
structure details may, on the one hand, be modified due to a change in the
\textit{c/a} ratio. On the other hand, nonuniform stresses in the basal plane
would result in a symmetry lowering which would split those bands that are
normally degenerate in the Brillouin zone AHL plane of hcp metals. In the case
of osmium such bands intersect the Fermi level, and a shear-stress-induced
splitting may thus have significant effect on the phonon self-energies.

In summary, by combining pressure- and temperature-dependent Raman
scattering experiments on osmium single crystals we have shown that
only a small fraction of the large temperature-induced frequency change
can be associated with the thermal expansion. Our results corroborate
the notion that both the large frequency variation and the decrease of
the phonon lifetime upon cooling originate from a non-adiabatic
electron-phonon interaction.

\acknowledgments

Yu.\ S. Ponosov acknowledges financial support from Deutscher Akademischer
Austauschdienst (DAAD).


\end{document}